\documentstyle[12pt,aasms4]{article}

\journalid{}{}
\articleid{}{}

\begin{document}

\title{BEAT CEPHEID PERIOD RATIOS FROM OPAL OPACITIES}

\author{Siobahn M. Morgan}
\affil{Department of Earth Science, University of Northern Iowa, \\
    Cedar Falls, IA 50614-0335, USA}
\and
\author{Douglas L. Welch}
\affil{Department of Physics and Astronomy, McMaster University, \\
    Hamilton, Ontario L8S 4M1, Canada}

\begin{abstract}
The discovery of a large number of beat Cepheids in the Large
Magellanic Cloud in the MACHO survey, 
provides an opportunity to compare the characteristics
of such Cepheids over a range of metallicities.  We produced a large grid
of linear nonadiabatic pulsation models using the OPAL opacity tables
and with compositions corresponding
to those of the Milky Way, and the Large and Small Magellanic Clouds.
Using the relationship between the period ratio and the main pulsation
period, we are able to define a range of models which correspond to
the observed beat Cepheids, and thereby constrain the 
physical characteristics of the LMC beat Cepheids.
We are also able to make some predictions about the nature of the
yet-to-be-discovered SMC beat Cepheids.
\end{abstract}


\section{Introduction}
Beat Cepheids (hereafter, BC's) provide us with valuable and precise
tests of stellar atmosphere and pulsation models. Indeed, the observed period
ratios in galactic BC's provided one of the most secure pieces of evidence
that the so-called `Cepheid mass discrepancy' --- the difference between
the masses inferred from models and those found from observations of Cepheids
in clusters or binary systems --- was real.

The sample of known BC's has grown recently as a result of
analysis of the photometric data collected by the MACHO Project in their
search for evidence of microlensing. \markcite{Alcock1995}Alcock 
{\it et al.} (1995) reported
45 new BC's in an analysis of variables in the 22 fields near the
bar of the Large Magellanic Cloud (LMC). 
(An {\it additional} 27 LMC BC's are now known).
This is in contrast to the 14 BC's currently known in the Milky Way.
The importance of this new sample arises from the expectation that
the LMC BC's are deficient in metals, compared to galactic BC's, by
a factor of 1.4--1.6 (\markcite{CC1986}Caldwell \& Coulson, 1986)
and the observational result that both the modal mix and period ratios
found for the LMC BC's are systematically different from their galactic
counterparts.

The sensitivity of the observable properties of Cepheids to metallicity
is still an open and important question with ramifications for both
our understanding of the evolution of intermediate-mass stars and for
the extragalactic distance scale. Period ratios, unlike most observationally
derived quantities, can be determined with very high precision -- 1 part
in 10$^5$ not being unusual. They are also direct measurements in the
sense that no intermediate relationship with other stars are assumed
and photometric calibration and reddening assumptions have no effect.
Obviously, a calibration of metallicity based on period ratio would
have great value. Such a relationship was seen by
\markcite{Aea1993}Andrievsky {\it et al.} (1993) who found a correlation
between [Fe/H] and $P_1$/$P_0$, where $P_0$ and $P_1$ are the periods
for the fundamental and first overtone modes, respectively. Comparison
with Figure 6 of \markcite{Aea1995}Alcock {\it et al.} (1995) suggests
that [Fe/H] can be predicted with even greater precision if it is a
function of both $\log P_0$ and $P_1$/$P_0$. In any case, the galactic
sample by itself is not well-suited to this investigation because of its
small number of stars and small range of metallicities.

\markcite{C-DP1995}Christensen-Dalsgaard \& Petersen (1995) used
existing M-L relations to fit pulsation
models to the BCs, but they were not able able to adequately match
the models to the $P_2$/$P_1$ pulsators in the LMC.  
\markcite{Buchleretal1996}Buchler {\it et al.} (1996) also examined
the characteristics of the LMC BCs to try to constrain the possible
masses of the stars.  In both of these studies, several assumptions were made,
particularly about the form of the M-L relation.

The motivation for this paper is to provide a set of pulsation model results,
based on current values for opacities, which can be used to interpret
both the galactic and LMC BC data in a single framework, and to
provide predictions for BC behavior in the yet-to-be-observed and even
more metal-deficient Small Magellanic Cloud (SMC) Cepheid sample.
We begin by describing our models, describe our results to date and
what we believe we have learned from the LMC sample, and conclude with
predictions for the SMC BC's.

\section{Pulsation Models}

Cepheid envelope models were analyzed for pulsational instability using a linear
nonadiabatic pulsation code, similar to that of \markcite{Castor1971}Castor 
(1971).  The stability of the fundamental mode as well as the 
first and second harmonics were examined.
For Cepheids in the Milky Way, abundances of
$X=0.7$, $Y=0.28$, and $Z=0.02$ were used.  The corresponding abundances for
the LMC and SMC Cepheids were ($0.7$, $0.29$, $0.01$) and ($0.7$, $0.296$, $0.004$),
respectively.  The models made use of the OPAL opacity tables using the
Grevesse and Noels solar composition mix 
(\markcite{IR1996}Iglesias \& Rogers, 1996).  Effective temperatures were chosen in the range 
5500 to 8400 K at increments of 100 K.  The envelopes for all models had
300 zones and were required to have a base zone temperature of at least $10^6$ K.
Even though convection has only a minimal influence on the pulsation periods
of the coolest models ($< 5800$ K), it was still included in the calculations
with a mixing-length ratio of 1.0 used throughout.

The mass ranges studied were varied according to sample, with the galactic 
Cepheid models encompassing  3 to 5.75 M$_{\sun}$, while the LMC
and SMC models had masses between 1.25 and 4.0 M$_{\sun}$. Rather than
using a single mass-luminosity relation, we selected a range of luminosity values
for each mass -- typically 5 different luminosities per mass.
The mass-luminosity ranges for the Milky Way Cepheid models cover the 
theoretically derived values obtained by 
\markcite{BIT1977}Becker, Iben \& Tuggle (1977),
the convective overshoot models of \markcite{Chiosi1990}Chiosi (1990), and the
Wesselink masses derived by \markcite{Simon1990}Simon (1990).  

In total, there were 1500 models of galactic Cepheids, and 1470 models
for the LMC and SMC Cepheids.  A sample of the results are given in
Tables 1 and 2 for the three compositions.  As expected, the lower metallicity
results in shorter periods and generally larger period ratios.
The blue edges for the instability
strips for each metallicity are given in Table 3.  The blue edges
were found to be hotter than those found by 
\markcite{CWC1993}Chiosi, Wood \& Capitano (1993), for LMC compositions.  In the case of
the fundmental mode models, the blue edge is approximately 
$\log T_{\hbox{eff}} = 0.07$ hotter, while the first overtone models are
hotter by $\log T_{\hbox{eff}} = 0.05$.  Typically, lower 
metallicities result in the blue edge moving to higher temperatures
for both the fundamental and first overtone mode pulsators.

Several relations for the models can be derived, such as the
period-mass-radius or `P-M-R' relation. For the models with metallicities
representative of the three different galaxies, we find:
\begin{eqnarray}
P_0 & = & 0.027 (M/M_{\sun})^{-0.66} (R/R_{\sun})^{1.68} \hbox{\ \ Milky Way,}\\
P_0 & = & 0.027 (M/M_{\sun})^{-0.65} (R/R_{\sun})^{1.67} \hbox{\ \ LMC, and} \\
P_0 & = & 0.025 (M/M_{\sun})^{-0.64} (R/R_{\sun})^{1.66} \hbox{\ \ SMC.}
\end{eqnarray}
These are very similar to the relations found by 
\markcite{Moskalik1995}Moskalik (1995) as well 
as those of \markcite{FSS1972}Fricke, Stobie, \& Strittmatter (1972).  

Another useful relation, and one which defines the parameter space
for each set of models is the period-luminosity-mass-effective
temperature or `P-L-M-T' relation. We find:

\begin{eqnarray}
\log (L/L_{\sun}) & = & -13.19 + 1.19 \log P_0 + 4.00 \log T_{\hbox{eff}} 
+ 0.78 \log (M/M_{\sun})  \hbox{\ \ Milky Way,} \\
\log (L/L_{\sun}) & = & -13.41 + 1.20 \log P_0 + 4.06 \log T_{\hbox{eff}}
+ 0.78 \log (M/M_{\sun})  \hbox{\ \ LMC, and} \\
\log (L/L_{\sun}) & = & -13.50 + 1.21 \log P_0 + 4.10 \log T_{\hbox{eff}}
+ 0.77 \log (M/M_{\sun})  \hbox{\ \ SMC.}
\end{eqnarray}
which are similar to those found by \markcite{Chiosi1990}Chiosi (1990), 
using Los Alamos opacities (\markcite{Hea1977}Huebner {\it et al.} 1977).

Petersen Diagrams ($\log P_0 - P_0/P_1$) for the models are shown in 
Figures 1, 2, and 3, as well as the location of observed BC's in the 
Milky Way and
the LMC.  The concentrations of models in the figures are due to 
identical luminosities, but varied masses.  Models near the
local maxima also tend to have similar values of effective temperature,
typically $T_{\hbox{eff}} = 5900 K$ for the Milky Way models and
$T_{\hbox{eff}} = 6100 K$ for the LMC models.   The temperature for
the Milky Way models is similar to that found for BCs by 
\markcite{Barrell81}Barrell (1981),
while the lower metallicity LMC models have both higher temperatures
and higher period ratios for identical masses and 
luminosities, as well as slightly shorter periods.  
In some cases the $P_2/P_1$ models extend into the region where 
the observed $P_1/P_0$ pulsators lie.  
While it may appear that there these stars could possibly be pulsating in
the $P_2/P_1$ mode, it should be noted that these are the lowest temperature
models 
($T_{\hbox{eff}} \approx 5500 K$),  and are much cooler than the
observed effective temperature for BC's.

\section{Analysis}

One distinction of the BC population in the LMC, compared to that in the Milky
Way, is the number of stars found at short periods ($\log P_0 < 0.4$).  Nearly
half of the LMC BC's are in this period range, as opposed to one galactic
Cepheid.  With such short periods, there are also
correspondingly lower luminosities and masses for these stars.  In the
set of LMC metallicity models discussed above, luminosities as low as 
$\log (L/L_{\sun}) = 2.25$ are needed to reproduce periods and period ratios
for these extremely short-period Cepheids.  The masses that correspond to
these models are typically less than $2 M_{\sun}$.  Larger masses
for comparable luminosities would result in lower values of the period
ratio.  

The pulsation models which fell closest to the location of the LMC BC's on the
Petersen diagram were from the LMC metallicity group.  The parameter
space these models covered corresponded to several different mass-luminosity
relations.

To determine which form of the mass-luminosity relation best describes
the LMC Cepheids,  further sets of models were produced using three 
different $M-L$ relations, 

\begin{eqnarray}
\log (L/L_{\sun}) & = & 3.43 \log M/M_{\sun} + 0.675 \\
\log (L/L_{\sun}) & = & 3.52 \log M/M_{\sun} + 1.0   \\
\log (L/L_{\sun}) & = & 1.64 \log M/M_{\sun} + 2.30 
\end{eqnarray}
where the first relation is from \markcite{BIT1977}Becker, 
Iben \& Tuggle (1977), 
using the composition of $X=0.7$, $Y=0.29$ and $Z=0.01$, the second is 
from \markcite{Chiosi1990}Chiosi (1990) for full convective overshooting, 
and the last is from \markcite{Simon1990}Simon (1990) based on a fit to
Wesselink masses for galactic Cepheids.  Models were produced to cover
the range of observed periods.  
The resulting Petersen diagrams are presented in 
Figures 4, 5, and 6 for both the F/1H and 1H/2H mode pulsators.  
For both the Becker, Iben \& Tuggle and Chiosi 
$M-L$ relations, the models are able to only fit the longest-period F/1H Cepheids.
Both $M-L$ relations fail to predict period ratios for the majority of the 
1H/2H pulsators. On the other hand, the Simon $M-L$ relation 
{\it does} reproduce the period ratios of most of the pulsators in
both groups.  The obvious difference is the slope of the
$M-L$ relation which is relatively small in the Simon relation.
As was pointed out by Christensen-Dalsgaard \& Petersen (1995) and 
Buchler {\it et al.} (1996), there are several aspects of LMC Cepheids
which are still puzzling.  In general, many types of evolutionary models
that are used to fit the characteristics of Cepheids in the LMC are
not always able to adequately satisfy all criteria.  The fact that
an $M-L$ relation, which was {\it not} based upon evolutionary models,
 appears to provide the best fit
for both the F/1H and 1H/2H BC's seems to further amplify this point.
This may indicate the need for further revisions in low metallicity
evolutionary models.

A second possible interpretation of the deviation between observations and theoretical
period ratios is that progressively lower metallicity BC's are found in the
instability strip as one goes to shorter periods (and, hence, lower masses
and luminosities).

\section{Predictions for the Beat Cepheids of the SMC}

Twelve field centers in the SMC are being monitored by the MACHO
Project. At this writing, these fields have not yet been searched
for variable stars. However, there are at least two indications 
that these fields contain additional BC's. 
\markcite{WS1965}Wesselink \& Shuttleworth (1965) reported ``a few
Small Magellanic Cloud Cepheids, with periods near three days, which
exhibit considerable irregularities in their light curves,'' which
they then suggested were the SMC counterparts of galactic BC's.
Second, \markcite{PGG1966}Payne-Gaposchkin \& Gaposchkin (1966) 
list numerous instances
of short-period Cepheids with unusually large scatter in their phased
lightcurves - a notation which \markcite{Alcock1995}Alcock {\it et al.} 
(1995) found correlated with BC activity in the similar catalogue of 
\markcite{PG1971}Payne-Gaposchkin (1971).

So far, we have used the observed period ratios in the Milky Way and
LMC BC's to determine the $M-L$ relation which matches the data for
the assumed metallicities. If our fits are correct, we can successfully
predict the period ratios for the even more metal-deficient SMC.
As before, we assume that the metallicity of the SMC is $X=0.7$, 
$Y=0.296$, and $Z=0.004$ for our models. 

The Petersen diagram of the SMC BC's is shown in 
Figure 3, which when compared to that found for the LMC composition
(Figure 2), reveals the familiar trend of systematically shorter periods
and higher period ratios for lower metallicities.  
If we assume that the BC's in the SMC have a similar mass, luminosity 
and temperature distribution as those in the LMC, we would find the
relationship between the period ratio and the fundamental mode period 
for the F/1H pulsators to be:

\begin{eqnarray}
P_1/P_0 & = & 0.740 - 0.033 \log P_0,      0.1 \leq \log P_0 \leq 0.69.
\end{eqnarray}
This relation assumes a similar slope as seen in the LMC, an
average increase in the F/1H period ratio of about 0.01, and a slight
decrease in the pulsation period.

The 1H/2H pulsators do not appear to be as sensitive to metallicity changes
as do the F/1H mode pulsators, so it is unlikely that their location on
the Petersen diagram would change significantly from one galaxy to another.

\section{Conclusions}

We find that the relatively shallow Simon mass-luminosity relation produces
pulsation models which most closely reproduce the MACHO 
Project BC period ratios 
reported in \markcite{Alcock1995}Alcock {\it et al.} (1995). We note the
possibility that the shortest-period BC's are systematically metal-poor
relative to their longer-period counterparts. We predict that the SMC
BC's will:

\begin{itemize}
\item{} display BC behavior to shorter periods and will consequently
be more numerous,
\item{} have higher period ratios for the F/1H stars than the LMC
by 0.01 on the average, and
\item{} have proportionally more 1H/2H pulsators due to the hotter
edge of the excitation region and the more extended blue loops typical
of the lower metallicity of the SMC Cepheids.
\end{itemize}

Further work will include the modeling of Cepheids over a greater range
of mass, luminosity and metallicity.  The models described in this
paper can be accessed via 
anonymous FTP at {\it nitro9.earth.uni.edu} (in {\it /pub/Cepheids })
and at the WWW site {\it http://nitro9.earth.uni.edu/Cepheids/ }.

\acknowledgments
SMM acknowledges the support of the Graduate College of the University of
Northern Iowa. This work was supported, in part, by a Research Grant from 
the Natural Sciences and Engineering Research Council of Canada (NSERC) to 
DLW.

\clearpage

%\plotone{morgan.fig1.eps}
\figcaption{Petersen diagram for galactic
composition models.  The lower sequence shows the location of the F/1H 
mode pulsators, while the upper sequence shows the dashed 1H/2H pulsators.  
The symbols indicate the different values for $\log (L/L_{\sun})$.
The progression of the models from the left to the right is due to the
increase in the primary period with decreasing temperature.
13 F/1H mode Beat Cepheids are shown
(filled circles), as well as a likely 1H/2H pulsator, CO Aur (open
circle).}

%\plotone{morgan.fig2.eps}
\figcaption{Petersen diagram for LMC composition models.  Symbols are
the same as in Figure 1.  LMC F/1H mode Beat 
Cepheids (Alcock {\it et al.}, 1995) are shown (filled
circles), as well as 1H/2H beat Cepheids (open circles). 
}

%\plotone{morgan.fig3.eps}
\figcaption{Petersen diagram for SMC composition models.  Symbols are the
same as in Figure 1.
}

%\plotone{morgan.fig4.eps}
\figcaption{Petersen diagram for LMC models using the 
Becker, Iben \& Tuggle (1977) $M-L$ relation.  The different masses for the
models are indicated by the symbols, with the F/1H mode pulsation models
on the lower region of the graph and the 1H/2H models on the upper region.
Also shown are the observed F/1H LMC beat Cepheids (large filled circles)
and the 1H/2H LMC beat Cepheids (large open circles).}

%\plotone{morgan.fig5.eps}
\figcaption{Petersen diagram for LMC models using the 
Chiosi (1990) $M-L$ relation.  
Symbols are the same as in Figure 4.} 

%\plotone{morgan.fig6.eps}
\figcaption{Petersen diagram for LMC models using the 
Simon (1990) $M-L$ relation.  
The different masses for the models are indicated by the symbols.
Also shown are the observed F/1H LMC beat Cepheids (large filled circles)
and the 1H/2H LMC beat Cepheids (large open circles).}

\clearpage
{\sc TABLE} 1. Fundamental Mode Models with $T_{\hbox{eff}} = 6000$ K.

{\sc TABLE} 2. First Harmonic Mode Models with $T_{\hbox{eff}} = 6000$ K.

{\sc TABLE} 3. Blue Edges for Milky Way, LMC and SMC Models, in units
of 100 K.

\end{document}